\documentclass[conference]{IEEEtran}

\usepackage{algorithmic}
\usepackage[cmex10]{amsmath}
\usepackage{graphicx}
\usepackage{amssymb,amsmath}
\usepackage{cite}
\usepackage{epsfig}
\usepackage{amsthm}

\usepackage[tight,footnotesize]{subfigure}
\usepackage{flushend}
\flushend

\newtheorem{definition}{\noindent\textbf{Definition}}

\newtheorem{remark}{\noindent\textbf{Remark}}
\newtheorem{theorem}{\noindent\textbf{Theorem}}
\newtheorem{lemma}[theorem]{\noindent\textbf{Lemma}}

\ifCLASSINFOpdf

\else

\fi

\begin{document}

\title{Constructing Sub-exponentially Large\\Optical Priority Queues with Switches and\\ Fiber Delay Lines}

\author{\IEEEauthorblockN{Bin~Tang, Xiaoliang~Wang, Cam-Tu Nguyen, Sanglu Lu}
\IEEEauthorblockA{National Key Laboratory for Novel Software Technology\\
 Nanjing University, Nanjing 210023, China\\
Email: \{tb, waxili, nguyenct,sanglu\}@nju.edu.cn\\
}}

\maketitle

\begin{abstract}
Optical switching has been considered as a natural choice to keep pace with growing fiber link capacity. One key research issue of all-optical switching is the design of optical queues by using optical crossbar switches and fiber delay lines (SDL). In this paper, we focus on the construction of an optical \emph{priority queue} with a single $(M+2)\times (M+2)$ crossbar switch and $M$ fiber delay lines, and evaluate it in terms of the buffer size of the priority queue. Currently, the best known upper bound of the buffer size is $O(2^M)$, while existing methods can only construct a priority queue with buffer $O(M^3)$.

In this paper, we make a great step towards closing the above huge gap. We propose a very efficient construction of priority queues with buffer $2^{\Theta(\sqrt{M})}$.
We use 4-to-1 multiplexers with different buffer sizes, which can be constructed efficiently with SDL, as intermediate building blocks to simplify the design. The key idea in our construction is to route each packet entering the switch to some group of four 4-to-1 multiplexers according to its current priority, which is shown to be collision-free.


%
%
%
%
%

\end{abstract}

\IEEEpeerreviewmaketitle

\section{Introduction}

All-optical packet switching is very attractive for making good use of the enormous bandwidth of optical networks since it eliminates the complicated and quite expensive optical-electronic-optical conversions. One main issue for implementing the all-optical packet switching is the construction of optical queues for conflict resolutions among packets competing for the same resources. As optical-RAM is not available yet, a common approach for constructing the optical queues is to use a combination of bufferless optical crossbar Switches and fiber Delay Lines (SDLs)~\cite{R03tso,H98slo,Y00aip}, where fiber delay lines act as storage devices for optical packets. However, fiber delay lines are much more inflexible than traditional electronic memories since one packet entering a fiber delay line can only be retrieved after a fixed amount of time. Such inflexibility makes the design of SDL-based optical queues with the same throughput and delay performance as that of the electronic ones quite challenging. In the past many years, great efforts have been made on constructing various kinds of optical queues, such as first-in-first-out (FIFO) multiplexers (e.g., \cite{C96cod,C07fsc,C96cor,C04rco,C06usd,C06ana,J07cof,J08oco}), FIFO queues (e.g., \cite{C06coo,L11mdq,H07rco,C13ana}), last-in-first-out (LIFO) queues (e.g., \cite{S07ams,H07rco,W11edo}), priority queues (e.g., \cite{S06eeo,C07asp,C07uas,K07oco}), shared queues (e.g., \cite{W09aco,W12cnt}), etc.

\begin{figure}[!tb]
    \centering
        \includegraphics[width=2.1in]{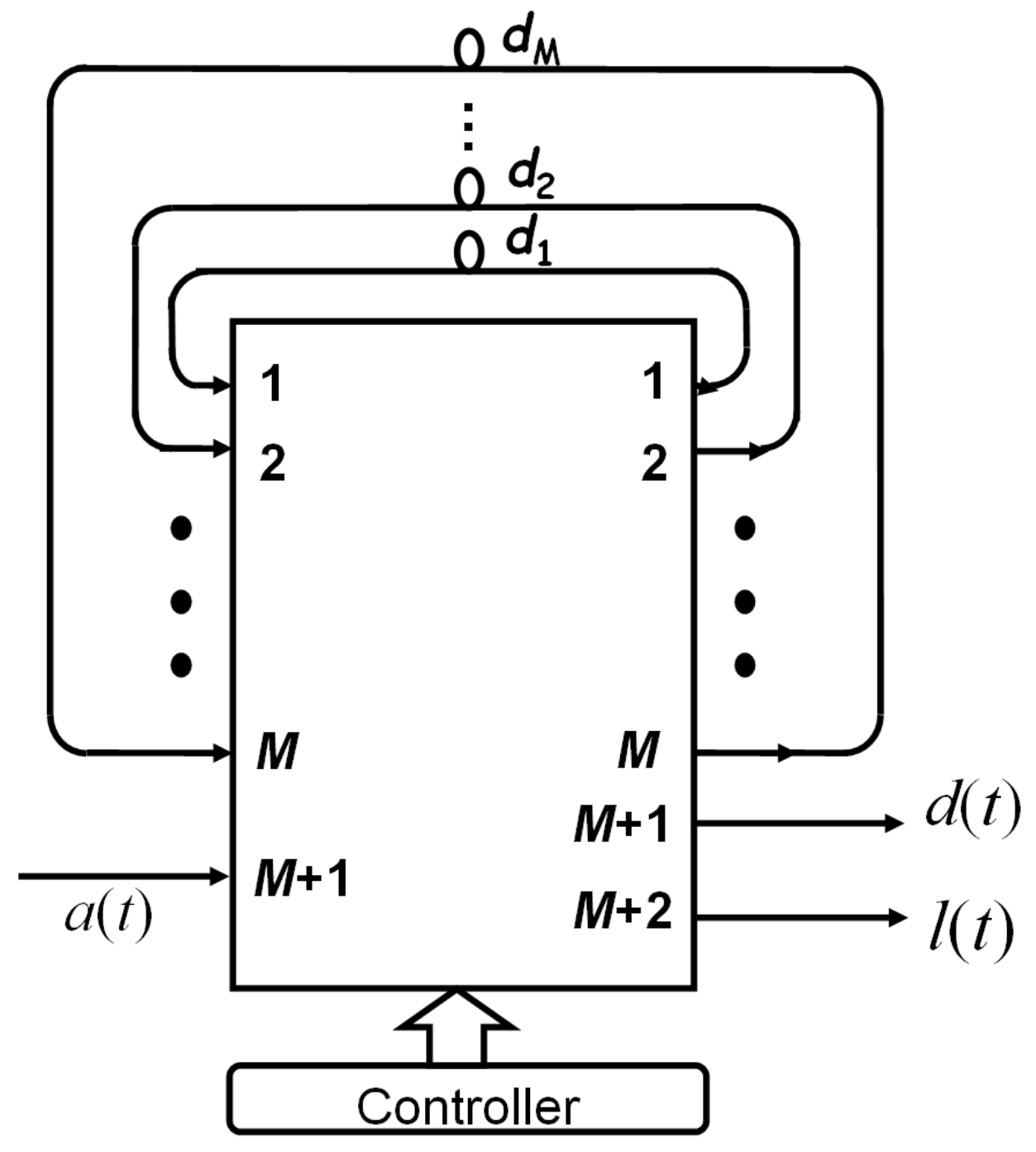}
    \caption{A construction of an optical priority queue with an $(M+2)\times (M+2)$ crossbar switch and $M$ fiber delay lines with delays $d_1$, \ldots $d_M$.}
    \label{fig:general}
\end{figure}

In this paper, we focus on the design of optical \emph{priority queues} with SDLs. In a priority queue, each packet is associated with a priority, and the packet with the highest priority is always sent to the output link when a departure request comes, whereas the packet with the lowest priority will be dropped when buffer overflow happens. Sarwate and Anatharam presented the first construction of optical priority queues in \cite{S06eeo}, where they considered a feedback system consisting of an $(M+2)\times (M+2)$ crossbar switch and $M$ fiber delay lines as illustrated in Fig.~\ref{fig:general}. Let $B^*$ be the maximum achievable buffer size of this feedback system acting as a priority queue. In \cite{S06eeo}, Sarwate and Anatharam showed that $B^*=O(2^M)$. They also gave an explicit construction of a priority queue using a sorting-based routing policy, which shows that $B^*=\Omega (M^2)$. A simpler proof for this result is given in \cite{C07asp}. Later, Chiu \emph{et al}.~\cite{C07uas} showed that $B^*=\Omega (M^3)$ by improving the Sarwate-Anatharam's sorting-based routing policy.
Despite the huge gap between the best known upper bound $O(2^M)$ and the best known lower bound $\Omega(M^3)$, little progress has been made on the bounds for almost ten years.

In this paper, we make a great step towards closing the above gap by giving a very efficient construction of optical priority queues, which leads to $B^*=2^{\Omega (\sqrt{M})}$. Instead of considering fiber delay lines as basic building elements, we use (FIFO) multiplexers with different buffer sizes as intermediate building blocks, which simplifies our design significantly. An $n$-to-1 multiplexer has $n$ input links and one departure link.  It always has a departure packet whenever it is nonempty and packets depart in the FIFO order. Our main idea is to let the packets with priorities currently belonging to a same properly chosen interval enter a corresponding group of multiplexers, and to introduce multiple multiplexers for each group and enough input links for each multiplexer to handle possible packet collisions. In particular, we show that using four 4-to-1 multiplexers in each group is enough for the system to be collision-free. Fortunately, a 4-to-1 multiplexer can be constructed by concatenating three 2-to-1 multiplexers with the same buffer size, whereas it has been proved in~\cite{C06ana} that an $(M+2)\times (M+2)$ crossbar switch and $M$ fiber delay lines are enough for constructing a 2-to-1 multiplexer with buffer $O(2^M)$. By further combining all the switches used into one, we finally have a construction of a priority queue with buffer $2^{\Theta (\sqrt{M})}$ using an $(M+2)\times (M+2)$ crossbar switch and $M$ fiber delay lines.
%

\section{Preliminaries}
\label{sec:preliminaries}
In this section, we first introduce the basic assumptions and network elements adopted in this paper and then introduce the definition of priority queues.

\subsection{Assumptions and Network Elements}

As in most work about the SDL-based optical queue designs~\cite{C06coo,C04rco,S06eeo,K07oco}, we assume that all packets have the same size, and time is slotted and synchronized so that every packet can be transmitted within one time slot. Since there is at most one packet in a link, we can use 0-1 variables to characterize the state of a link. We say that a link is in state 1 at time $t$ if there is a packet in the link at time $t$, and the link is in state 0 at time $t$ otherwise.

Switches and fiber delay lines are defined as follows.
\begin{definition}[\textbf{Switch}]
  An $M\times M$ (optical) crossbar switch is a network element that has $M$ input links and $M$ output links, which can realize all the $M!$ permutations between its inputs and outputs.
\end{definition}
\begin{definition}[\textbf{Fiber delay line}]
A fiber delay line with delay $d$ (a non-negative integer) is a network element that has one input link and one output link, through which $d$ time slots are required for a packet to traverse.
\end{definition}

\subsection{Priority Queues}

\begin{figure}
    \centering
        \includegraphics[width=2.7in]{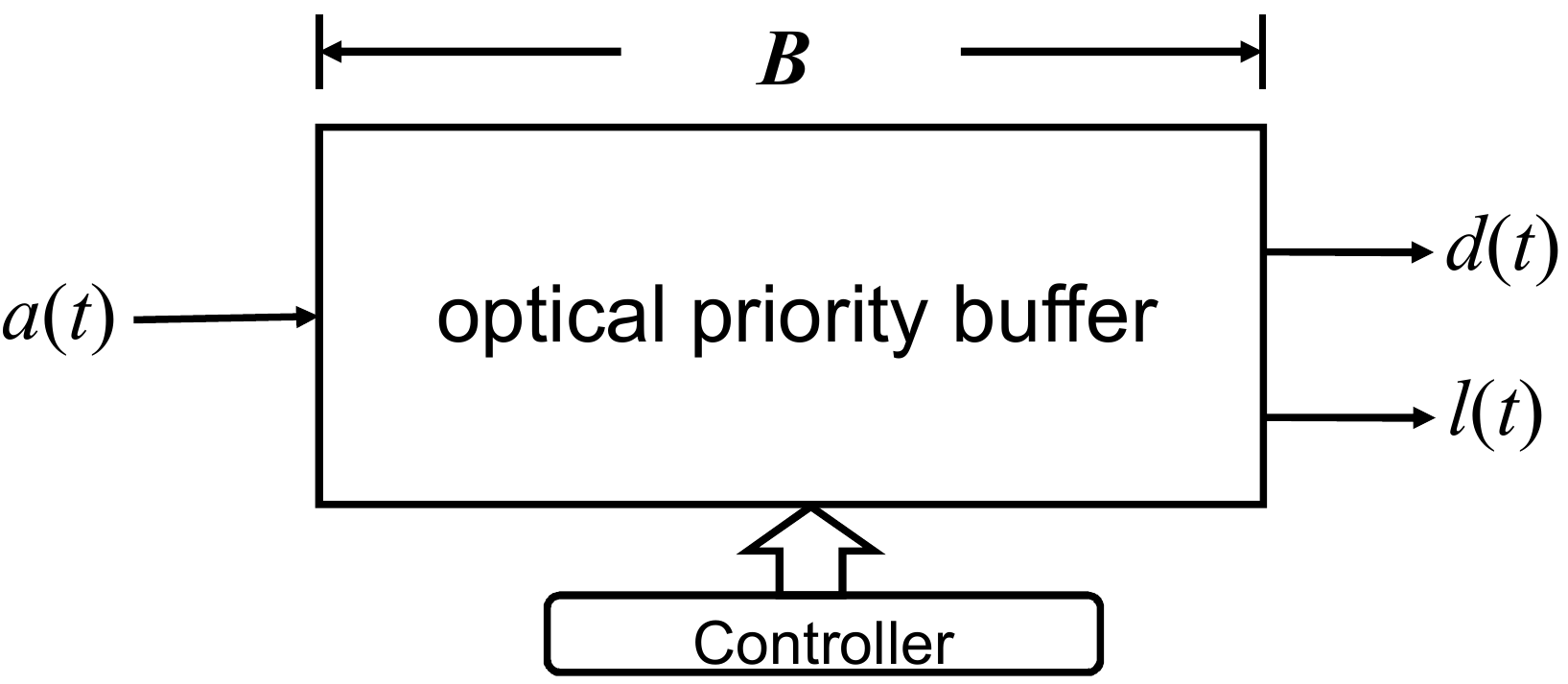}
    \caption{An optical priority queue with buffer size $B$}
    \label{fig:priorityQ}
\end{figure}

In this paper, we focus on the design of optical priority queue with switches and fiber delay lines. A priority queue with buffer $B$ is a network element with certain properties, which has one input link, one controller, and two output links, one for departing packets and the other for loss packets. See Fig.~\ref{fig:priorityQ} for an illustration. To formally define the properties of a priority queue, we first introduce some basic notations describing the state of the priority queue at each time $t$ (i.e., the $t$-th time slot).
\begin{itemize}
  \item Let $a(t)$, $d(t)$ and $l(t)$ denote the states of the input link, the departure link and the loss link at time $t$, respectively.
%
  \item Let $c(t)=1$ if the controller sends a departure request at time $t$ or $c(t)=0$ otherwise.
  \item Denote by $q(t)$ the number of packets in the queue at time $t$.
\end{itemize}

\begin{definition}[\textbf{Priority}]
When a packet arrives at the queue, it is assigned with a priority, which indicates the expected departure order of this packet among all the buffered packets. Let $i$ be an arriving packet or a packet buffered in the queue at time $t$. Let $r_i(t)$ denote the priority order of $i$ among all the packets buffered in the queue and the arriving packet (if any) at time $t$, i.e., if $i$ has the $j$-th highest priority, then $r_i(t)=j$.
\end{definition}

Based on the aforementioned notations, a discrete-time priority queue can be defined as follows.

\begin{definition}[Priority queue~\cite{S06eeo}]
\label{def:sharequeue}
The network element shown in Fig.\ref{fig:priorityQ} is called a priority queue with buffer $B$ if it satisfies the following properties:
\begin{enumerate}
\item[(P1)] \emph{Flow conservation}: arriving packets are either stored in the queue or transmitted through the departure link or the loss link, i.e.,
\begin{equation}
  q(t)=q(t-1)+a(t)-d(t)-l(t).
\end{equation}

\item[(P2)] \emph{Non-idling}: There is always a packet departing from the queue if and only if the controller sends a departure request, and there are packets in the queue or there is an arriving packet. In other words,
    \begin{equation}
      d(t)=
      \begin{cases}
        1 & \textrm{if }c(t)=1 \textrm{ and }q(t-1)+a(t)>0\\
        0 & \textrm{otherwise.}
      \end{cases}
    \end{equation}

\item[(P3)] \emph{Maximum buffer usage}: There is a lost packet if and only if there is no departure request, the buffer is full and there is an arriving packet, i.e.,
\begin{equation}
  l(t)=
  \begin{cases}
    1 & \textrm{if }c(t)=0, q(t-1)=B \textrm{ and }a(t)=1\\
    0 & \textrm{otherwise.}
  \end{cases}
\end{equation}
Also arriving packets are dropped only when the queue is full, i.e.,
    \begin{eqnarray}
     l(t)= [q(t-1)+a(t)-d(t)-B]^{+}
    \end{eqnarray}
where $[x]^{+}=x$ if $x > 0$ and is otherwise 0.
\item[(P4)] \emph{Priority departure}: If there is a departure packet $i$ at time $t$, then $i$ has the highest priority among all the packets buffered in the queue and the arriving packet (if any) at time $t$, i.e.,
    \begin{equation}
      r_i(t)=1.
    \end{equation}

\item[(P5)] \emph{Priority loss}: If there is a loss packet $i$ at time $t$, then $i$ has the lowest priority among the packets buffered in the queue and the arriving packet at time $t$, i.e.,
    \begin{equation}
      r_i(t)=B+1.
    \end{equation}
\end{enumerate}
\end{definition}


\section{A Construction of Priority Queues}
\label{sec:construction}

In this section, we present a very efficient construction of priority queues where 4-to-1 FIFO multiplexers are used as intermediate building blocks.

\subsection{Multiplexers}
\begin{definition}[\textbf{Multiplexer}~\cite{C04rco}]
  An $n$-to-1 (FIFO) multiplexer with buffer $B$ is a network element with $n$ input links and $n$ output links. Among the $n$ output links, one is for departing packets and the others are for lost packets. Let $a_i(t)$, $i=1,2,\ldots,n$, be the state of the $i$-th input link, $d(t)$ be the state of the departure link and $l_i(t)$, $i=1,2,\ldots,n-1$, be the state of the $i$-th loss link, and $q(t)$ be the number of packets buffered at the multiplexer at time $t$. The $n$-to-1 multiplexer with buffer $B$ satisfies the following four properties.
  \begin{itemize}
    \item [(M1)] Flow conservation: arriving packets from the $n$ input packets are either stored in the buffer or transmitted through the $n$ input links, i.e.,
    \begin{equation}
      q(t)=q(t-1)+\sum_{i=1}^na_i(t)-d(t)+\sum_{i=1}^n l_i(t).
    \end{equation}

    \item [(M2)] Non-idling: there is always a departing packet if there are packets in the buffer or there are arriving packets, i.e.,
    \begin{equation}
      d(t)=\begin{cases}
        1 & \textrm{if } q(t-1)+\sum_{i=1}^n a_i(t)>0\\
        0 & \textrm{otherwise.}
      \end{cases}
    \end{equation}

    \item [(M3)] Maximum buffer usage: arriving packets are lost only when the buffer is full, i.e., for $i=1,\ldots,n-1$,
    \begin{equation}
      l_i(t)=\begin{cases}
        1 & \textrm{if }q(t-1)+\sum_{i=1}^na_i(t)\geq B+i+1\\
        0 & \textrm{otherwise.}
      \end{cases}
    \end{equation}

    \item [(M4)] FIFO: packets depart in the FIFO order.

  \end{itemize}
\end{definition}

\begin{figure}[!tb]
    \centering
        \includegraphics[width=2.7in]{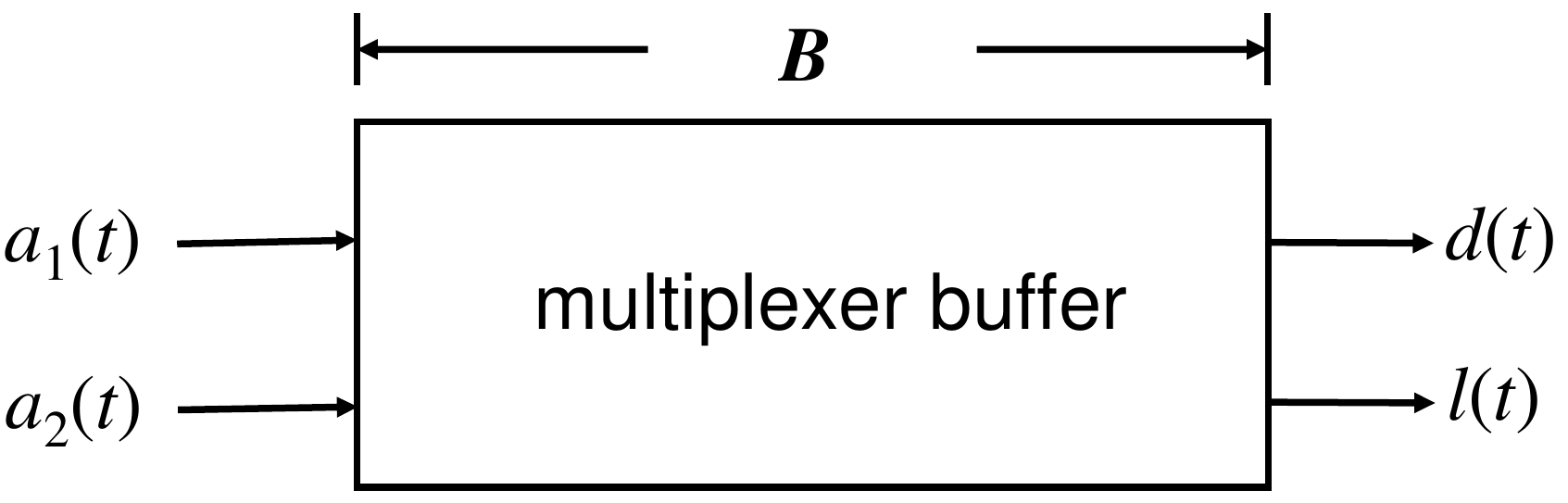}
    \caption{A 2-to-1 multiplexer with buffer $B$.}
    \label{fig:2to1}
\end{figure}

See Fig.~\ref{fig:2to1} for an illustration of a 2-to-1 multiplexer with buffer $B$.

\subsection{The Construction}

To ease the presentation, we first define some notations. Let $m$ be a positive integer. For $j=1,\ldots,m$, let $\Psi_j$ be the set of consecutive integers $2^{j-1},2^{j-1}+1,,\ldots, 2^j-1$, and for $j=m+1,\ldots,2m-1$, let $\Psi_j$ be the set of consecutive integers  $3\times 2^{m-1}-2^{2m-j}, \ldots, 3\times 2^{m-1}-2^{2m-j-1}-1$. For example, $\Psi_1=\{1\}$, $\Psi_2=\{2,3\}$, and $\Psi_3=\{4,5,6,7\}$. Clearly, for each $j=1,\ldots,m$, both $\Psi_j$ and $\Psi_{2m-j}$ have the same size of $2^{j-1}$.

\begin{figure*}
    \centering
        \includegraphics[width=5.2in]{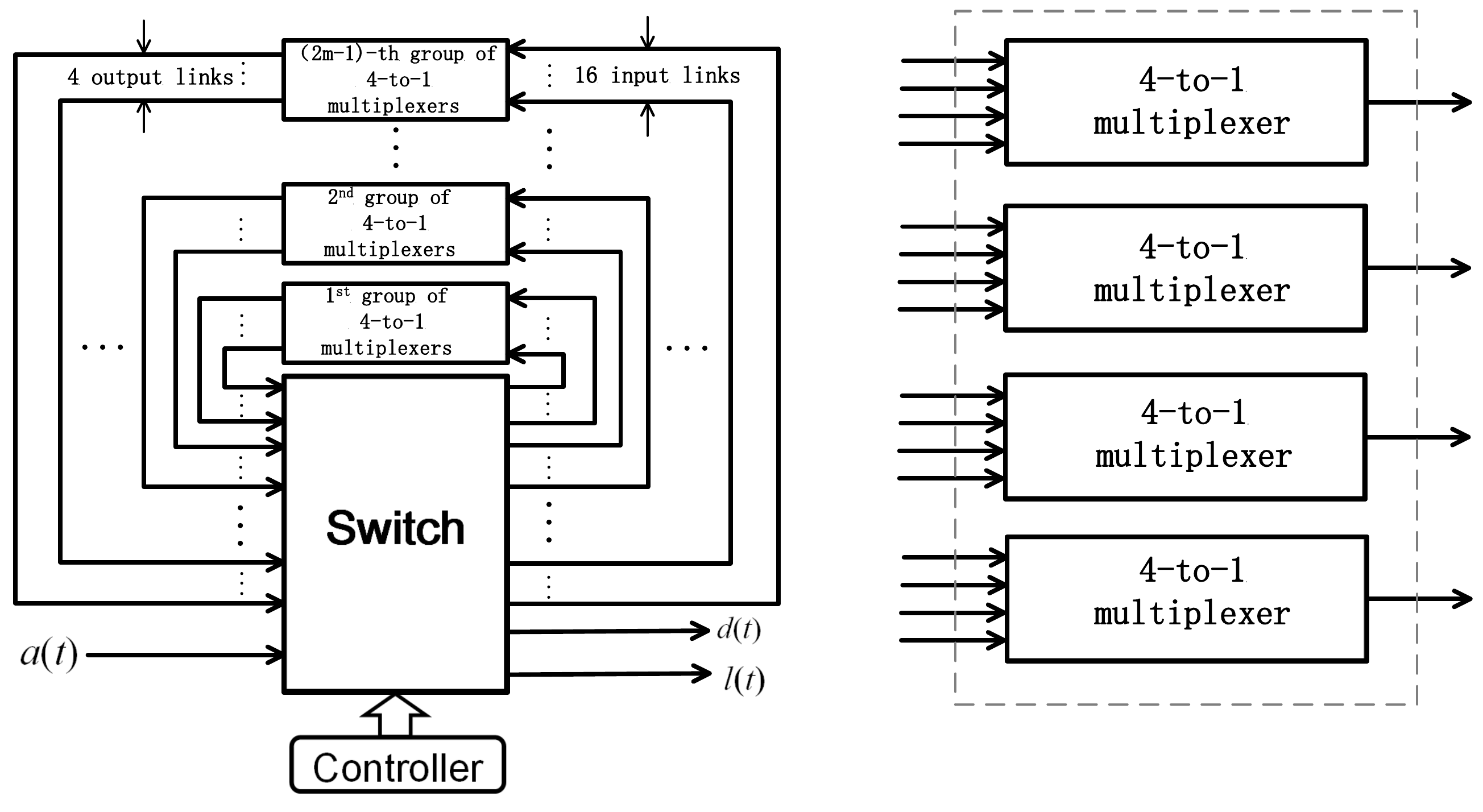}
    \caption{Left: the construction of an optical priority queue with a switch and $2m-1$ groups of 4-to-1 multiplexers. Right: A group of 4-to-1 multiplexers. All the loss links of multiplexers are omitted.}
    \label{fig:construction}
\end{figure*}

In our construction, an optical priority queue, as illustrated in Fig.~\ref{fig:construction}, consists of a $(32m-14)\times (32m-14)$ crossbar switch and $2m-1$ groups of multiplexers. For each $j=1,2,\ldots,m$, the $j$-th group of multiplexers consists of four parallel 4-to-1 multiplexers with buffer $B_j$, where
\begin{equation}
  B_j=
  \begin{cases}
    1 & i=1\\
    2^{j-2} & i=2,3,\ldots,m,
  \end{cases}
  \end{equation}
and its main purpose is to store the packets with priority order in $\Psi_j$, although the priority orders of packets may change slowly over time. The $(2m-j)$-th group of multiplexers is the same as the $j$-th group of multiplexers, and its main purpose is to store the packets with priority order in $\Psi_{2m-j}$. The construction of 4-to-1 multiplexers with switches and fiber delay lines is deferred to Sec.~\ref{sec:4to1}.

The routing policy performed by the switch at the beginning of time $t$, $t=1,2,\ldots,$ is as follows.
 \begin{itemize}
 \item[(R)] After handling the departure packet and the loss packet, if any, according to the requirement of priority queues, every other packet $i$ in the switch with priority order $r_i(t)$ will be pushed into one of the 4-to-1 multiplexers in the $j$-th group such that
\begin{equation}
\label{eq:enterrule}
  r_i(t)\in \Psi_j,
\end{equation}
via a separate input link while guaranteeing that all the four buffers of the 4-to-1 multiplexers in this group are almost the same used (i.e., differing by at most one packet).
\end{itemize}
For example, the packets with priority order 4,5,6,7 are always sent to the third group of multiplexers.

%
%

\subsection{Analysis}

Now we show that the system constructed above emulates a priority queue with buffer $3\times 2^{m-1}-2$ exactly.

We first establish the correctness of the routing policy by showing that there is no packet collision under the routing policy, i.e., the number of packets entering every group of 4-to-1 multiplexers at the same time is not more than 16, the total number of input links of the group of 4-to-1 multiplexers. We begin with the following property.
\begin{lemma}
  \label{lemma:multiplexinterval}
    For any packet $i$ buffered in the $j$-th group of multiplexers at time $t$, if $j\leq m$, then
  \begin{equation}
  \label{eq:multiplexinterval}
    2^{j-2}+1\leq r_i(t) \leq 2^j+2^{j-2}-2,
  \end{equation}
  otherwise,
  \begin{equation}
    3\times 2^{m-1}-5\times 2^{2m-j-2}+1\leq r_i(t)\leq 3\times 2^{m-1}-2^{2m-j-2}-2.
  \end{equation}
\end{lemma}
\begin{IEEEproof}
  We only prove the result for the case $j\leq m$. The result for the other case can be shown very similarly.

    Suppose that $j\leq m$ and a packet $i$ is buffered in some 4-to-1 multiplexer in the $j$-th group at time $t$. Let $t_0\leq t$ be the time when $i$ entered this multiplexer for the last time.  During the time interval between $t_0$ and $t$, since the multiplexer is non-empty, there is always a packet departing from the multiplexer in the FIFO order according to (M2) and (M4). So we have $t-t_0\leq B_j-1$. Otherwise, $i$ should leave the multiplexer before $t$. Another key observation is that the priority order of any packet can only change by one in a time slot. This implies that
  \begin{equation}
    |r_i(t)-r_i(t_0)|\leq t-t_0\leq B_j-1.
  \end{equation}
  Since $2^{j-1}\leq r_i(t_0)\leq 2^j-1$ according to the routing policy (R), we have  \eqref{eq:multiplexinterval} immediately.
\end{IEEEproof}

\begin{lemma}
\label{thm:nocollision}
  No collision can happen under the proposed routing policy (R).
\end{lemma}
\begin{IEEEproof}
  According to \eqref{eq:enterrule} and Lemma~\ref{lemma:multiplexinterval}, it is straightforward to check that every packet entering the $j$-th group of multiplexers can only come from the input link of the switching system, the output links of the $(j-1)$-th group of multiplexers, the output links of the $j$-th group of multiplexers, or the output links of the $(j+1)$-th group of multiplexers, if any. So there are at most 13 packets entering the $j$-th group of multiplexers at the same time under the routing policy (R). Clearly, there is no collision as the $j$-th group of multiplexers has 16 input links in total.
\end{IEEEproof}

Now we establish the main result.

\begin{theorem}
\label{thm:main}
  Starting with an empty buffer, the proposed switching system is an optical priority queue with buffer $3\times 2^{m-1}-2$.
\end{theorem}
\begin{IEEEproof}
  We first show that the proposed system satisfies the flow conservation property (P1). According to Theorem~\ref{thm:nocollision}, we only need to show that no packet is dropped by any 4-to-1 multiplexer in the system. Consider an arbitrary $j$-th group of multiplexers such that $2\leq j\leq m$ (the case $j=1$ is trivial while the case $j>m$ can be shown similarly). Since the buffers of the four 4-to-1 multiplexers are almost equally used, according to the routing policy (R), if there is a packet dropped by some 4-to-1 multiplexer in the group, then all the four buffers must be full at that time. However, this could not happen as the four buffers have a total size of $4\times B_j=2^j$ but there are at most $2^j-2$ packets that can be stored in this group according to Lemma~\ref{lemma:multiplexinterval}. Therefore, (P1) always holds.

  Now we prove this result by induction. Suppose that the proposed system emulates the priority queue with buffer $3\times 2^{m-1}-2$ up to time $t$. Trivially, this holds for $t=0$. We will show that it also emulates the priority queue with buffer $3\times 2^{m-1}-2$ at time $t+1$.

  We first show that (P2) and (P4) hold at time $t+1$.  Without loss of generality, we suppose that $c(t+1)=1$. By the induction hypothesis, we can assume that there is a departure packet at time $t$. If there is only one packet in the system at time $t$, then (P2) and (P4) hold at time $t+1$ directly. Otherwise, there is a packet $i$ such that $r_i(t)=2$. If there is an arriving packet with higher priority than $i$ at time $t+1$, then the packet will depart immediately according to the routing policy, so (P2) and (P4) are satisfied. Otherwise, $r_i(t+1)=1$. According to Lemma~\ref{lemma:multiplexinterval}, $i$ must be stored in the second group or the third group of multiplexers at time $t$. For the former case, $i$ will depart from the second group at $t+1$ and thus depart from the switching system according to (R). For the latter case, $i$ will also depart from the third group at $t+1$. Otherwise, $i$ should have a priority order of at most 3 at the time it entered the third group for the last time since $B_3=2$, which contradicts with the routing policy (R). In all cases, we have shown that (P2) and (P4) hold at time $t+1$. By conducting a similar argument, we can show that (P3) and (P5) also hold at time $t+1$. The proof is accomplished.
\end{IEEEproof}

\subsection{Design of 4-to-1 Multiplexers and The Complexity}
\label{sec:4to1}

The design cost of our construction depends on how to construct 4-to-1 multiplexers with SDL. As demonstrated in the above analysis, some requirements of the used 4-to-1 multiplexers could be relaxed. First, any 4-to-1 multiplexer with buffer $B_j$ could be replaced by a 4-to-1 multiplexer with buffer larger than or equal to $B_j$. Second, as no packet would be dropped by any 4-to-1 multiplexer, we may use \emph{simplified 4-to-1 multiplexer} instead, for which it is unnecessary to satisfy the property (M3) strictly.
\begin{figure}[!tb]
    \centering
        \includegraphics[width=2.6in]{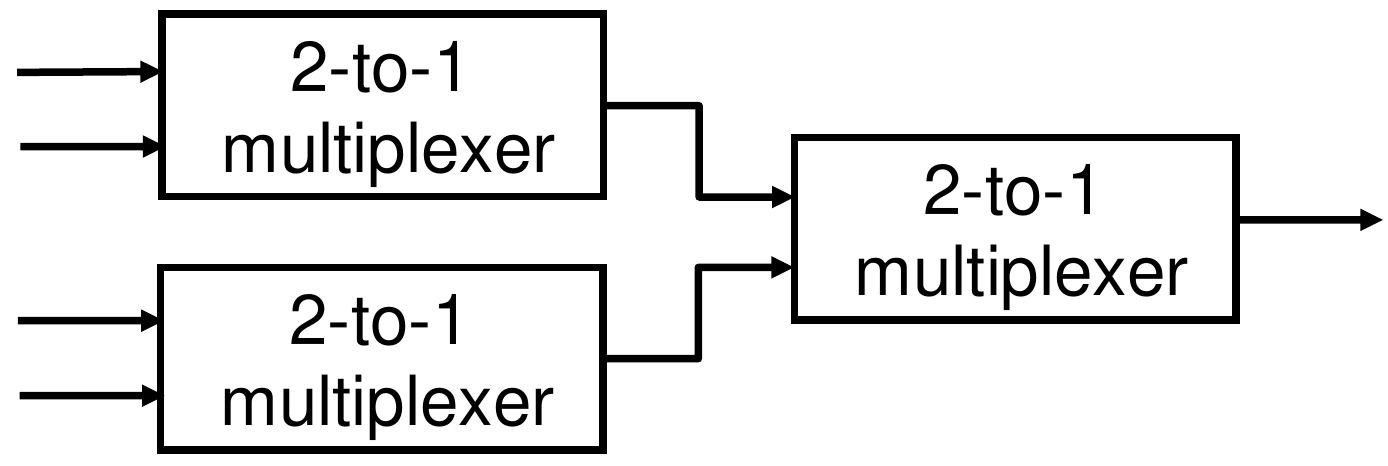}
    \caption{A simplified  4-to-1 multiplexer. Here we omit all the loss links.}
    \label{fig:4t01}
\end{figure}
A simplified 4-to-1 multiplexer with buffer $B$ can be constructed by concatenating three 2-to-1 multiplexers with buffer $B$.  See Fig.~\ref{fig:4t01} for an illustration. Due to the space limitation, we omit the details here.  In \cite{C06ana}, Chou \emph{et al.} proposed an efficient construction of 2-to-1 multiplexers with buffer at least $B$ with an $(M+2)\times (M+2)$ crossbar switch and $M$ fiber delay lines, where $M=\lceil \log_2(B+1) \rceil$. Based on this, we can have the following construction cost.
\begin{theorem}
  A priority queue with buffer size $B=3\times 2^{m-1}-2$ can be constructed with a $(32m-14)\times (32m-14)$ crossbar switch, forty-eight $3\times 3$ switches, twenty-four $(j+1)\times (j+1)$ switches for each $j=3,\ldots,m-1$, twelve $(m+1)\times (m+1)$ switches, and 12($m^2-2m+3)$ fiber delay lines.
\end{theorem}
Despite we use many switches in our construction, one can combine all these switches into one for the possible reduction of the hardware cost. We thus have the following theorem.
\begin{theorem}
  A priority queue with buffer $B=3\times 2^{m-1}-2$ can be constructed with a $(12m^2+56m-2)\times (12m^2+56m-2)$ crossbar switch and $12(m^2-2m+3)$ fiber delay lines.
\end{theorem}

In other words, a priority queue with buffer size $B$ can be constructed with an $(M+2)\times (M+2)$ crossbar switch and $M$ fiber delay lines where $M=O(\log^2B)$, or equivalently, $B=2^{\Omega(\sqrt{M})}$.

\begin{remark}
The construction cost of our priority queues can be reduced by, e.g., replacing the first/last group of multiplexers by a single fiber delay line, or replacing one of the 4-to-1 multiplexers in each group by a 2-to-1 multiplexer with the same buffer size (see the proof of Lemma~\ref{thm:nocollision}), but our result will not change in the order sense.
\end{remark}

\bibliographystyle{IEEEtran}
\bibliography{IEEEabrv,reference}



%
%
%

\end{document}